\documentclass[journal]{new-aiaa}
\usepackage[utf8]{inputenc}

\usepackage{graphicx}
\usepackage{amsmath}
\usepackage[version=4]{mhchem}
\usepackage{siunitx}
\usepackage{longtable,tabularx}
\usepackage{epsfig}
\usepackage{xcolor}
\usepackage{subcaption}
\usepackage{epstopdf}

\newcommand{\norm}[1]{\left\Vert#1\right\Vert} 
\newcommand{\pd}[2]{{\frac{\partial #1}{\partial #2}}} 
\newcommand{\vect}{\underrightarrow}
\DeclareMathAlphabet{\mbf}{OT1}{ptm}{b}{n}
\newcommand{\mbs}[1]{{\boldsymbol{#1}}}
\newcommand{\mbc}[1]{ \boldsymbol{\mathcal{#1}} } 
\newcommand{\ura}[1]{{\underrightarrow{{#1}}}}
\newcommand{\trans}{{\ensuremath{\mathsf{T}}}} 
\newcommand{\beq}{\begin{equation}}
\newcommand{\eeq}{\end{equation}}

\title{Areostationary Satellite Station Keeping Via a Natural Motion Trajectory and Predictive Control\footnote{A version of this note~\cite{gall2025predictive} (AIAA 2025-99106) was presented at the AIAA Region V Student Conference, April 3-4, 2025 in Minneapolis, MN.}}

\author{Nathan A. Gall\footnote{Ph.D. Student, Department of Aerospace Engineering and Mechanics, University of Minnesota, 110 Union St. SE, Minneapolis, MN 55455, AIAA Student Member}, Robert D. Halverson\footnote{Ph.D. Candidate, Department of Aerospace Engineering and Mechanics, University of Minnesota, 110 Union St. SE, Minneapolis, MN 55455, AIAA Student Member} and Ryan J. Caverly\footnote{Associate Professor, Department of Aerospace Engineering and Mechanics, University of Minnesota, 110 Union St. SE, Minneapolis, MN 55455, AIAA Member.}}
\affil{University of Minnesota - Twin Cities, Minneapolis, MN, 55455}

\begin{document}

\maketitle

\section{Introduction}
\lettrine{A}{chieving} a sustained human presence on Mars will require satellite coverage to facilitate communication and navigation. At the Earth, geostationary satellites play a key role in terrestrial communication and navigation constellations, as they provide continuous coverage over a wide area. Satellites in areostationary Mars orbits (AMO), the Martian equivalent to a geostationary Earth orbit (GEO), have been proposed to fulfill a similar role to their Earth-orbiting counterparts~\cite{Hastrup2003,Lock2016,Edwards2016,Babuscia2017,Breidenthal2018,pontani2022mars}. When considering only the two-body problem, an AMO satellite will maintain its longitude without the need for any station keeping. In reality, satellites in AMO experience non-Keplerian perturbing forces that cause them to drift from their nominal orbit~\cite{Romero2017,matthieu2025areostationary}. A similar effect is present in GEO, where gravitational perturbations induce a significant change in the orbit's inclination (North-South direction) and a slow precession in the East-West direction~\cite{soop1994handbook}. Station keeping is performed periodically using thrusters and on-board propellant to correct for these deviations. These station-keeping maneuvers can be scheduled manually from the ground or performed autonomously on-board the satellite, using high-impulse chemical propulsion or low-thrust propulsion~\cite{deBruijn2016,gazzino2019long,Caverly2020,de2025combined}. It is imperative that station-keeping maneuvers be performed in an efficient manner, as the lifetime of a satellite is dictated by its fuel remaining on board. The amount of fuel required for station keeping is computed in terms of $\Delta v$, which is defined in this note as the $L_1$ norm of thrust normalized by the satellite's mass.

The perturbations experienced by AMO satellites are notably different than those present in GEO. In particular, the perturbations in the orbit's inclination (North-South direction) are an order of magnitude smaller, while the East-West and radial perturbations are much more significant~\cite{Silva2013,Romero2015}.  This results in relatively quick deviations from a nominal areostationary orbit, which -- along with the considerably large communication delay between Earth-based ground stations and Martian satellites -- motivate the need for autonomous AMO station keeping control strategies.

 Model predictive control (MPC) has been investigated for use with many spacecraft applications~\cite{eren2017model,di2018real,petersen2023safe}, including autonomous GEO station keeping~\cite{Caverly2020}. MPC directly optimizes control actions through a finite length of time in the future, called the prediction horizon. The optimal control sequence is recomputed at each time step, resulting in a feedback policy. In order to directly optimize over the prediction horizon, a model of system dynamics is used to forecast how control actions affect the system's trajectory. There is an inherent tradeoff between complexity in the prediction model and computational effort of the MPC policy. Developing a prediction model with linear system dynamics is desirable since it enables MPC to be solved through well-developed and efficient convex optimization strategies, which is critical when considering autonomous implementation on-board a satellite. MPC has been investigated for AMO station keeping~\cite{Halverson2021,Halverson2023,gall2024,Halverson2025}, where it was shown that a significant tradeoff exists between fuel-efficiency and computational-efficiency. Specifically, the use of nonlinear MPC resulted in an annual $\Delta v$ of $3.8$~m/s, while requiring computation power well-beyond the capabilities of flight computers~\cite{Halverson2023,Halverson2025}. An MPC policy using linear time-invariant (LTI) dynamics resulted in an annual $\Delta v$ of $4.5$~m/s and was realistically solvable on-board a satellite as a quadratic program~\cite{Halverson2021}. A linear time-varying (LTV) dynamic model was shown to yield a computationally-efficient MPC policy with an annual $\Delta v$ of $3.7$~m/s~\cite{gall2024}, although the study in Ref.~\cite{gall2024} noted that MPC policies with LTI or LTV dynamics exhibited performance that was very sensitive to the tuning of the MPC policy objective function and prediction horizon length.

This note presents a novel approach to MPC-based AMO station keeping that bridges the gap seen in prior work between fuel-efficiency and computational-efficiency. This is accomplished through the discovery and use of a fuel-free natural motion trajectory in the presence of the dominant radial and East-West non-Keplerian perturbing forces that maintains the satellite within one degree of longitude from a areostationary orbit. Two of these natural motion trajectories exist as limit cycles about the stable equilibrium longitudes located at $17.92^\circ$ West and $167.83^\circ$ East. They are the resulting motion in the presence of Mars' non-homogeneous gravitational field, accounting for Keplerian and higher-order gravitational perturbations. The proposed MPC policy uses an LTV dynamic model that is derived by linearizing the satellite's dynamics relative to the appropriate natural motion trajectory. The result is a station keeping policy that minimizes the fuel consumed, maintains thrust and station-keeping constraints, and is computationally tractable for on-board implementation as a quadratic program.

The novel contributions of this note include 1) an MPC-policy that, to the best of the knowledge of the authors', achieves the lowest $\Delta v$ of any autonomous AMO station keeping policy that is formulated as a convex optimization problem and 2) a thorough assessment of the proposed MPC policy's robustness to realistic model uncertainty, computation time requirements, and estimation errors. The robustness study presented in this note extends far beyond previous studies on MPC-based AMO station keeping~\cite{Halverson2021,Halverson2023,gall2024,Halverson2025}, which all assumed perfect model information, instantaneous computations, and perfect state knowledge.

\section{Preliminaries}

This section defines relevant vector notation, important reference frames, and the direction cosine matrix used to describe the relative orientation of the reference frames.

\subsection{Notation}
Matrices are presented in boldface. Physical vectors are denoted as $\vect{(\cdot)}$. A reference frame $\mathcal{F}_c$ is defined by three orthonormal basis vectors $\ura{a}^1$, $\ura{a}^2$, and $\ura{a}^3$, arranged in a right-handed fashion. The position vector describing the location of point $b$ relative to point $a$ is $\vect{r}^{ba}$, with its components resolved in reference frame $\mathcal{F}_c$ written as $\mbf{r}^{ba}_c = [r^{ba}_{c1} \,\, r^{ba}_{c2} \,\, r^{ba}_{c3}]^\trans$.

\subsection{Reference Frames}

The center of Mars is defined by the unforced point $w$. The reference frame $\mathcal{F}_{a}$ is defined such that $\vect{a}_1$ extends from $w$ to $0^\circ$ latitude and longitude at January 1st, 2000 noon GMT (commonly referred to as the J2000 epoch). The basis vector $\vect{a}_3$ extends from $w$ in the direction of geographical North, while $\vect{a}^2$ is chosen to complete the definition of the right-handed reference frame. This frame is non-rotating and referred to as the Mars Centered Inertial (MCI) frame. A nominal areostationary orbit is described by a point $h$ such that $\vect{r}^{hw}$ is located in the equatorial plane and passes through a specified longitude $\lambda$. Assuming Mars rotates at its mean rotation rate $n$ corresponding to approximately one rotation every $24.6$ hours, $\|\vect{r}^{hw}\|_2$ must be $2.04277 \times 10^4$ km to be consistent with a relative two-body orbit. This distance is defined as $r_\text{nom}$; the nominal AMO semi-major axis. A second reference frame, $\mathcal{F}_{h}$, is defined such that $\vect{h}^1$ lies collinear with $\vect{r}^{hw}$, and $\vect{h}^3$ lies collinear with $\vect{a}^3$. This frame, referred to as Hill's frame, is related to the MCI frame by the direction cosine matrix (DCM) $\mbf{C}_{ha}(t) = \mbf{C}_3(n(t-t_\text{J2000})+\lambda)$, where $\mbf{C}_3(\cdot)$ is the DCM associated with a principle rotation about $\vect{a}^3$ and $t_\text{J2000}$ is the time of the J2000 epoch.

\section{Dynamics of a Perturbed Areostationary Satellite}\label{dynamics}

This section presents the dynamics of a satellite in a perturbed areostationary orbit, beginning with its general equations of motion, followed by a description of the relevant perturbations, a natural motion trajectory of the satellite, and a discrete-time, time-varying model describing its motion relative to the natural motion trajectory.

\subsection{Equations of Motion}

Consider a spacecraft, modeled as a point with mass $m_B$, located at point c in orbit around Mars. The state of the satellite, $\mbf{x}(t) \in \mathbb{R}^6$, at time $t$ is taken to be its position relative to a nominal areostationary orbit, resolved in Hill's frame, and the rate of change of that quantity, which is defined as $ \mbf{x}(t) = [ \mbf{r}^{ch^\trans}_h(t) \,\, \dot{\mbf{r}}^{ch^\trans}_h(t) ]^\trans$.  The dependence of state, control, and perturbations on time are omitted throughout this note, except when necessary for clarity. It is assumed that the spacecraft is equipped with the ability to thrust in each axis. The equations of motion for this system are defined by the motion of the satellite relative to an unperturbed circular reference areostationary orbit and are given by~\cite{Alfriend2010}
\begin{equation}\label{eq:NLDynamics}
    \hspace{-10pt}\dot{\mbf{x}}
    = \mbf{f}(\mbf{x},\mbf{u},\mbf{w}(\mbf{x},t)) = \left[\begin{matrix} 
   \dot{\mbf{r}}^{ch}_h \vspace{5pt} \\
    2n\dot{r}^{ch}_{h2} + n^2r^{ch}_{h1} - \frac{\mu(r_\text{nom} + r^{ch}_{h1})}{\left[(r_\text{nom}+r^{ch}_{h1})^2+(r^{ch}_{h2})^2 + (r^{ch}_{h3})^2\right]^{3/2}}+ \frac{\mu}{r_\text{nom}^2} + w_{h1}(\mbf{x},t) + \frac{u_1}{m_B} \vspace{5pt} \\
    -2n\dot{r}^{ch}_{h1} + n^2r^{ch}_{h2} - \frac{\mu r^{ch}_{h2}}{\left[(r_\text{nom}+r^{ch}_{h1})^2+(r^{ch}_{h2})^2 + (r^{ch}_{h3})^2\right]^{3/2}} + w_{h2}(\mbf{x},t) + \frac{u_2}{m_B} \vspace{5pt} \\
    -\frac{\mu r^{ch}_{h3}}{\left[(r_\text{nom}+r^{ch}_{h1})^2+(r^{ch}_{h2})^2 + (r^{ch}_{h3})^2\right]^{3/2}} + w_{h3}(\mbf{x},t) + \frac{u_3}{m_B}
    \end{matrix}\right],
\end{equation}
where $\mu$ is the Martian gravitational constant, $\mbf{u} = [ u_{h1} \,\, u_{h2} \,\, u_{h3} ]^\trans$ is the thrust vector, expressed in Hill's frame, and $\mbf{w}(\mbf{x} , t) = [w_{h1}(\mbf{x} , t) \,\, w_{h2}(\mbf{x} , t) \,\, w_{h3}(\mbf{x} , t)]^\trans $ is the acceleration experienced by the spacecraft due to exogenous perturbations, also expressed in Hill's frame.

\subsection{Perturbations}
The exogenous forces experienced by the spacecraft that are considered in this analysis fall into three categories: third-body gravity $\vect{a}_\text{3rd}$, solar radiation pressure $\vect{a}_\text{SRP}$, and gravitational harmonics $\vect{a}_\text{harm}$. The exogenous perturbation force is therefore decomposed as
\begin{equation}\label{allpert}
    \vect{w} = \vect{a}_\text{3rd}+\vect{a}_\text{SRP}+\vect{a}_\text{harm}.
\end{equation}
Third-body gravity considers the gravitational force on the spacecraft from a body that is not Mars and takes on the form
\begin{equation}
    \vect{a}_\text{3rd} = \sum_i -\frac{\mu_i \vect{r^{ci}}}{\norm{\vect{r}^{ci}}^{3}},
\end{equation}
where $\mu_i$ is the gravitational parameter of the $i^\text{th}$ body, and $\vect{r}^{ci}$ is the position of the spacecraft relative to the $i^\text{th}$ body. For the simulations in this study, the effects of Mars' moons Phobos and Deimos and the Sun are considered. 

The effects of solar radiation pressure are small, and a discussion of their mechanics is beyond the scope of this note. A discussion of their effects on AMO station keeping is presented in Ref.~\cite{Halverson2025}.

The most significant perturbation experienced by areostationary satellites is caused by Mars' nonuniform gravitational potential. The orbits of terrestrial satellites are influenced greatly by Earth's equatorial bulge, termed the J2 perturbation. J2 refers to the term a particular series expansion describing a primary body's gravitational potential~\cite{Chao2018-ho}. The basis functions for this expansion are called spherical harmonics. 
Mars' gravitational potential can similarly be described using spherical harmonic expansion. The gravitational potential $U$, is given in spherical coordinates $r,\lambda,\phi$: radial distance, longitude, and latitude respectively, and is described by~\cite{Chao2018-ho}
\begin{equation}\label{eq:gravPot}
  U = \frac{\mu}{r}\left[\sum_{l=0}^{\infty}\left(\frac{R_M}{r} \right)^l \sum_{m=0}^{l} \left(C_{lm}\cos{m\lambda}+S_{lm}\sin{m\lambda}\right)P_{lm}(\sin{\phi})\right],
\end{equation}
where $R_M$ is Mars' equatorial radius, $C_{lm},S_{lm}$ are coefficients, and $P_{lm}$ are Legendre polynomials. The perturbation due to the gravitational anomaly is given by
\begin{equation}\label{eq:gravPert}
    \vect{a}_\text{harm} = \nabla U - \frac{\mu \vect{r}^{cw}}{\norm{\vect{r}^{cw}}^3},
\end{equation}
where the nominal Keplerian gravitation term has been subtracted to leave only the perturbing acceleration. In this note, the coefficients from the GMM2B model presented in Ref.~\cite{Lemoine2001} are used up to a fifth order. This is consistent with the modeling choice used in Ref.~\cite{Halverson2025}, where it was shown that higher-order terms beyond fifth order result in a negligible change in the resulting gravitational acceleration.

The East-West (i.e., `along-track') gravitational anomaly perturbations experienced by an areostationary satellite are heavily dependent on longitude. There exist two longitudes that are stable equilibria for East-West motion; $17.92^\circ$ West and $167.83^\circ$ East~\cite{Alvarellos2009-hc}. These longitudes---denoted ``stable longitudes''---provide attractive targets for areostationary missions, since they not only experience small perturbations in one principle axis, but they also experience a restoring force when deviating from the nominal longitude. 

\subsection{Natural Motion Trajectory}
\label{sec:NaturalMotionTraj}
\begin{figure}[t!]
    \centering
    \includegraphics[width=0.49\linewidth]{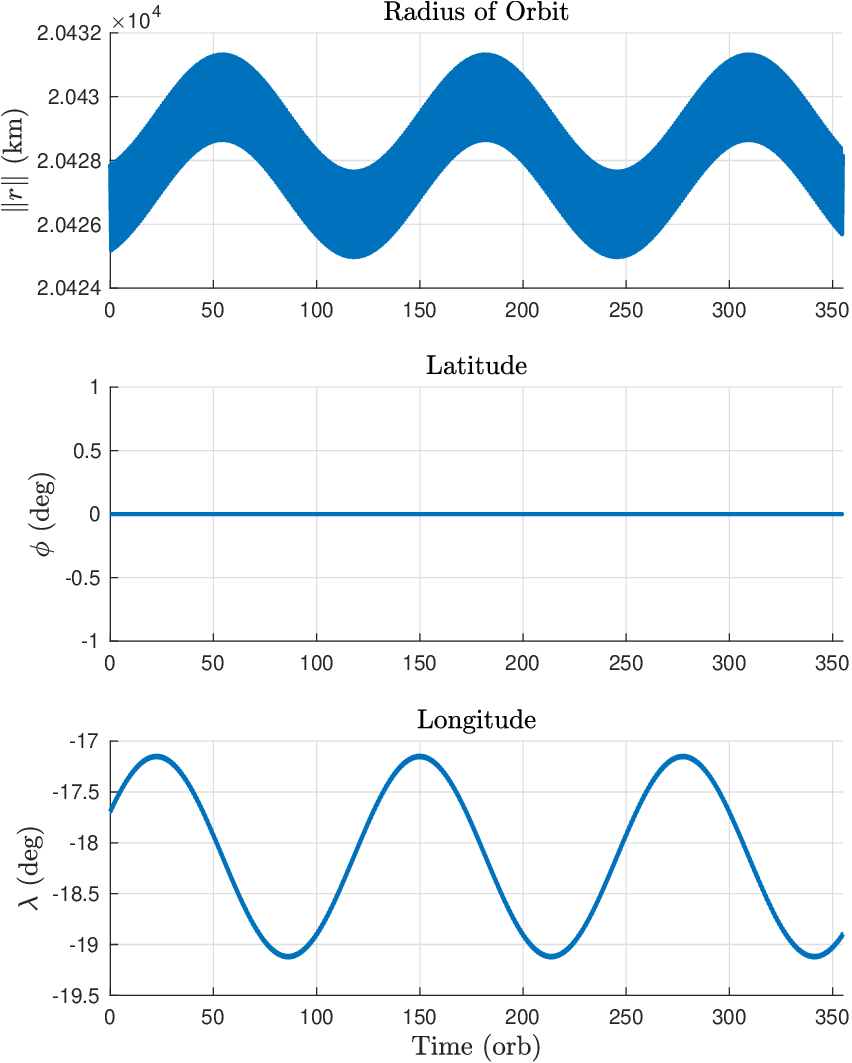}
    \caption{The natural motion trajectory caused by spherical harmonic perturbations about $17.92^\circ$ West.}
    \label{fig:LimitCycle}
\end{figure}

Consider a satellite in AMO at a stable longitude of $17.92^\circ$ West. The perturbations experienced by the satellite due to the gravitational anomaly modeled to a fifth order are shown in Table~\ref{tab:AMO perts}. The radial perturbation is nearly three orders of magnitude greater than the perturbations in the other two directions at this longitude. The coupling between East-West and radial motion seen in Eq.~\eqref{eq:NLDynamics} combined with the restorative force in the East-West perturbation induce a periodic variation in the satellite's orbit in the form of a limit cycle. This motion is shown in Fig.~\ref{fig:LimitCycle}, where the orbit periodically falls ahead and behind of the nominal AMO. This manifests as a drift in longitude by about a degree in either direction with a period of approximately 127 days.  Since this bounded longitudinal motion is caused by the radial and along-track spherical harmonics, a satellite following this trajectory would theoretically only have to account for the North-South (cross-track) spherical harmonic and non-spherical-harmonic perturbations to remain on this trajectory. Allowing for this drift effectively eliminates the need to cancel out the effects of the largest perturbation during station keeping. Trigonometric analysis shows that if a receiver is located on the Martian surface at $17.92^\circ$ West, a deviation in an areostationary satellite's longitude of $1^\circ$ would correspond to a $1.2^\circ$ change in elevation from the horizon.

\begin{table}[t!]
    \centering
            \caption{Spherical harmonic perturbations experienced by an AMO satellite at longitude $17.92^\circ$ West.}
    \begin{tabular}{cc}
        \hline \hline
        Axis & Acceleration (m/s$^2$)\\
        \hline Radial & $-6.7448\times10^{-9}$\\
        East-West & $7.9315\times10^{-12}$\\
        North-South & $-2.0131\times10^{-11}$\\
        \hline \hline
        \end{tabular}
\label{tab:AMO perts}
\end{table}

A station keeping policy proposed in this note attempts to keep the satellite state close to the state trajectory defined by this limit cycle. The satellite's state on this trajectory is given by $ \bar{\mbf{x}}(t)$, which evolves according to 
\begin{equation}\label{eq:cycleDynamics}
    \dot{\bar{\mbf{x}}} = \mbf{f}(\bar{\mbf{x}},\mbf{0},\bar{\mbf{w}}(\bar{\mbf{x}})),
\end{equation}
where $\bar{\mbf{w}}$ contains only the radial and along track terms of the perturbations.

\subsection{Discrete-Time Linear Time-Varying Model}\label{model}
This section describes a linear prediction model approximating the nonlinear dynamics in Eq.~\eqref{eq:NLDynamics} in a similar to the method presented in Ref.~\cite{gall2024}. The goal is to generate a discrete-time, linear time-varying (LTV) model that approximates the motion of the point-mass satellite relative to a reference trajectory, from each temporal node to the next. This replaces the nonlinear system of differential equations describing the dynamics with an approximate system of linear algebraic equations, which is computationally tractable to use as a prediction model in real time. In this case, the limit cycle state $\bar{\mbf{x}}$ is chosen as the reference state trajectory. Since the limit cycle is a natural motion trajectory, the reference control trajectory is defined as $\bar{\mbf{u}} = \mbf{0}$.

A continuous-time LTV system is constructed by taking a first-order Taylor's series expansion about a reference trajectory, given by
\beq \label{eq:ltvcont}
\dot{\mbf{x}}(t) \approx \mbf{A}(t)\delta\mbf{x}(t) + \mbf{B}(t)\delta\mbf{u}(t) + \mbf{z}(t),
\eeq 
where $\delta\mbf{x} = \mbf{x}-\bar{\mbf{x}}$, $\delta\mbf{u} = \mbf{u}-\bar{\mbf{u}}$, $\mbf{A}(t) = \pd{}{x}\mbf{f}(\mbf{x},\mbf{u},\mbf{w},t)\left.\right\vert_{\bar{\mbf{x}},\bar{\mbf{u}},\mbf{w}(\bar{\mbf{x}}),t}$, $ \mbf{B}(t) = \pd{}{u}\mbf{g}(\mbf{x},\mbf{u},t)\left.\right\vert_{\bar{\mbf{x}},\bar{\mbf{u}},\mbf{w}(\bar{\mbf{x}}),t}$,  and $\mbf{z}(t) = \mbf{f}(\bar{\mbf{x}},\bar{\mbf{u}},\mbf{w}(\bar{\mbf{x}}),t)$. In Eq.~\eqref{eq:ltvcont}, the perturbations $\mbf{w}(\bar{\mbf{x}})$ are evaluated at the reference state, but include the terms neglected in $\bar{\mbf{w}}$, allowing the resulting prediction model to capture the cross-track terms.

For the prediction model to be used within a direct numerical optimization framework, it must express the dynamics between discrete timesteps. Applying a zeroth-order hold~\cite{Malyuta2019} to the inputs, the discrete-time LTV system dynamics are given by
\beq\label{eq:ltvdisc}
\mbf{x}_{k+1} \approx \mbf{A}_k\mbf{x}_k+\mbf{B}_k\mbf{u}_k+\mbf{z}_k,
\eeq
where $\mbf{x}_k$ and $\mbf{u}_k$ are the state and control input at $t_k$, respectively, and
\begin{subequations}\label{eq:disccomponents}
\begin{alignat}{3}
\mbf{A}_k &= \mbs{\Phi}(t_{k+1},t_k),\\
\mbf{B}_k &= \mbf{A}_k \int_{t_k}^{t_{k+1}}\mbs{\Phi}(\tau,t_k)^{-1}\mbf{B}(\tau)\mathrm{d}\tau,\\
\mbf{z}_k &=\int_{t_k}^{t_{k+1}}\mbs{\Phi}(\tau,t_k)^{-1}\mbf{z}(\tau)\mathrm{d}\tau.
\end{alignat}
\end{subequations}
The state transition matrix $\mbs{\Phi}(t_{k+1},t_k)$ is obtained by numerically integrating $ \frac{\mathrm{d}}{\mathrm{d}t}\mbs{\Phi}(t,t_0) = \mbf{A}(t)\mbs{\Phi}(t,t_0)$ subject to the initial condition $\mbs{\Phi}(t_0,t_0) = \mbf{1}$. As $\mbf{A}_k,\mbf{B}_k,$ and $\mbf{z}_k$ are found through integration of values of $\mbs{\Phi}$ between time indices, Eq.~\eqref{eq:disccomponents} and the state transition matrix can be numerically integrated simultaneously.

This formulation is quite general, and implementation must be catered to the specific AMO station-keeping application. In this system, the calculation of $\mbf{B}(t)$ is trivial and can be done analytically. The calculation of $\mbf{A}(t)$ is complicated by the spatial dependence of the disturbance terms. In Ref.~\cite{gall2024}, the nominal areostationary orbit was used as the reference trajectory where the spatial dependence of the perturbations are insignificant, thus $\pd{\mbf{w}}{x}$ was approximated as $\mbf{0}$. In this note, the limit cycle is used as a reference, where the trajectory strays hundreds of kilometers from the nominal stationary point, causing this approximation to no longer be accurate. As the partial derivatives of truncations of Eq.~\eqref{eq:gravPot} are cumbersome to write analytically, $\mbf{A}(t)$ is calculated using a complex-step numerical differentiation scheme inspired by Refs.~\cite{martins2003complex,cao2025complex}, which makes use of the function \texttt{atan2ComplexStep.m}~\cite{ricciardi2025complex} in MATLAB.

In the implementation of control policies, it is often advantageous to define states relative to the desired reference, resulting in the perturbed dynamics
\begin{equation}\label{eq:relsystem}
    \delta\mbf{x}_{k+1} \approx \mbf{A}_k\delta\mbf{x}_k+\mbf{B}_k\delta\mbf{u}_k+\delta\mbf{z}_k,
\end{equation}
where
\begin{equation}\label{eq:dz}
\delta\mbf{z}_k = \mbf{A}_k\bar{\mbf{x}}_k+\mbf{B}_k\bar{\mbf{u}}_k + \mbf{z}_k - \bar{\mbf{x}}_{k+1}.
\end{equation}
The term $\delta \mbf{z}_k$ accounts for the difference between the reference dynamics and a truly open-loop trajectory, a consequence of the cross-track perturbation terms being absent in $\bar{\mbf{w}}$. This completes the derivation of the discrete-time LTV prediction model linearized about a natural motion trajectory.

\section{Model Predictive Control Station-Keeping Methodology}

The MPC station-keeping policy proposed in this work is similar to the linear time-varying MPC policy presented in Ref.~\cite{gall2024}, with the main difference being that the dynamics are defined relative to the natural motion trajectory described in Section~\ref{sec:NaturalMotionTraj}. At each time step, the MPC policy optimizes control actions a set number of time steps in the future, called the prediction horizon. The objective of this policy is to minimize a quadratic function of state deviation and control action across the prediction horizon using a linear time-varying discrete-time model to predict how the states evolve across the horizon. The first time step of the optimal control input sequence is then applied until the next time step, when the algorithm repeats. 
It should be noted that the propagation of the system to obtain the model in Eq.~\eqref{eq:ltvdisc} is a fully open-loop process, and thus may be computed a priori. The key difference in this novel implementation compared to the method in Ref.~\cite{gall2024} is that rather than using the AMO orbit as the reference trajectory, the dynamics used in the prediction model in Eq.~\eqref{eq:relsystem} are evaluated along the limit cycle discussed in Section~\ref{dynamics}.  Previous policies achieved fuel savings by allowing a satellite to drift within a defined station keeping window near the desired latitude and longitude. By redefining the states relative to the limit cycle in this work, the control policy allows for similar drift relative to the longitudinal variation in the natural motion trajectory. The MPC problem is formulated as
\begin{align*}\label{eq:LQ-MPCcost}
& \underset{\mbc{X}_t,\mbc{U}_t,\mbs{\nu}}{\text{minimize}}
& & \mbs{\nu}^\trans \mbf{S} \mbs{\nu} + \sum^{N-1}_{k=0}\delta\mbf{u}_{k|t}^\trans \mbf{R} \delta\mbf{u}_{k|t}\\
& \text{subject to} \\
& & &\delta\mbf{x}_{k+1|t} = \mbf{A}_{k|t}\delta\mbf{x}_{k|t}+\mbf{B}_{k|t}\delta\mbf{u}_{k|t}+\delta\mbf{z}_{k|t}, \quad  \forall k \in \left[0 \quad \cdots \quad N-1\right], \\
& & &\delta\mbf{x}_{0|t} = \delta\mbf{x}(t), \\
& & &\delta\mbf{x}_\text{min} - \mbs{\nu} \leq \delta\mbf{x}_{k|t} \leq \delta\mbf{x}_\text{max} + \mbs{\nu}, \quad  \forall k \in \left[0 \quad \cdots \quad N\right], \\
& & &\mbf{f}_\text{min}^\text{thrust} \leq \delta\mbf{u}_{k|t} \leq \mbf{f}_\text{max}^\text{thrust}, \quad \forall k \in \left[0\quad \cdots \quad N-1\right], \\
& & &\mbf{0} \leq \mbs{\nu},
\end{align*}
where $N$ is the prediction horizon, $\mbc{X}_t = \{\delta\mbf{x}_{0|t},\ldots,\delta\mbf{x}_{N-1|t}\}$, $\mbc{U}_t = \{\delta\mbf{u}_{0|t},\ldots,\delta\mbf{u}_{N-1|t}\}$, and $\mbf{R} = \mbf{R}^\trans > 0$ is the constant control weighting matrix. The subscript notation $k|t$ denotes the state or control input $k$ prediction steps ahead of time $t$. For example, $\mbf{x}_{k|t}$ is the predicted state $k$ steps ahead of time $t$. 
The variables $\delta\mbf{x}_\text{min}$ and $\delta\mbf{x}_\text{max}$ are defined based on the prescribed station keeping window in which the spacecraft can drift from the natural motion trajectory. Specifically, $\delta\mbf{x}_\text{max} = [ \infty \,\, r_\text{nom} \tan(\lambda_\text{max}) \,\, r_\text{nom} \tan(\phi_\text{max}) \,\, 0 \,\, 0 \,\, 0 ]^\trans$ and $\delta\mbf{x}_\text{min} = -\delta\mbf{x}_\text{max}$, where $\lambda_\text{max}$ and $\phi_\text{max}$ are the maximum allowable drift in longitude and latitude, respectively. The slack variable $\mbs{\nu}\in \mathbb{R}^n$ relaxes the station-keeping window constraint and allows for infeasible initial conditions. This is referred to as a ``soft constraint." Without this, the solution to the optimization problem would not exist should the spacecraft ever leave the window. With a sufficiently large weight $\mbf{S} = \mbf{S}^\trans > 0$, the controller strictly enforces the window while the spacecraft is on the interior. If model inaccuracies or unforeseen environmental factors lead to the spacecraft exiting the window, the single slack variable across the prediction horizon will prevent further drift outside the window, while discouraging over-actuation to satisfy the bounds. 
The maximum and minimum allowable thrust inputs expressed in Hill's frame are defined as $\mbf{f}_\text{max}$ and $\mbf{f}_\text{min}$, respectively, and $\mbf{f}_\text{min} = -\mbf{f}_\text{max}$.

\section{Numerical Results}\label{results}

Numerical simulation results are presented in this section to demonstrate the performance of the proposed station-keeping policy. A description of the simulation setup is first presented, followed by nominal results with the proposed station-keeping policy, comparisons to existing AMO station-keeping policies in the literature, and a robustness study of the proposed station-keeping policy.

\subsection{Simulation Setup}
\label{sec:SimSetup}

The spacecraft is approximated as a point mass with mass of $m_B = 4000$~kg, solar facing area of $37.5$~m$^2$, and a solar radiation constant of $4.5\times10^{-6}$. The spacecraft is placed in a nominal AMO at a stable longitude of $17.92^\circ$ West. The spacecraft's maximum thrust is set to $\mbf{f}_\text{max} = [50 \,\, 50 \,\, 50 ]^\trans$~mN based on the assumption of low-thrust propulsion.  The station-keeping window is defined by maximum longitude and latitude deviations of $\lambda_\text{max} = 0.2^\circ$ and $\phi_\text{max} = 0.05^\circ$.  For simulations with the proposed policy, these maximum deviations are defined relative to the natural motion trajectory, while simulations of the policies in Refs.~\cite{Halverson2021,Halverson2023,Halverson2025,gall2024,gall2025predictive} constrain these longitude and latitude deviations relative to the true AMO stable longitude of $17.92^\circ$ West and $0^\circ$ of latitude.

The MPC station-keeping policy uses discrete time steps of length $\Delta t = 1$~hour and a prediction horizon of $18$~hours ($N = 18$) was chosen through tuning of the proposed policy to ensure satisfactory performance across all results. Values of $\mbf{R} =5.6 \times  10^5 \cdot \mbf{1}$, and $\mbf{S} = 100 \cdot \mbf{1}$ 
are used within the cost function. These variables are chosen to emphasize the desire to minimize control effort, without concern as to where the spacecraft is within the station-keeping window. This MPC optimization problem is expressed as a convex quadratic problem, and solved using MATLAB's \texttt{quadprog.m} function, using the \texttt{interior-point-convex} algorithm.

All numerical simulations presented in this work begin on January 1st, Noon GMT, 2000 and involve running the appropriate station-keeping policy for $570$ orbits.  The first $215$ orbits to allow the North-South station keeping bound to be reached (i.e., the system is undergoing a transient response), while the last $355$ orbits represent one Earth year of steady-state response.  All annual $\Delta v$ computations are made over the final $355$ orbits to be representative of the steady-state fuel requirements. To remain consistent with the zero-order hold used in the model of the MPC policy, the control thrust is held constant between time steps. The system is propagated forward in time according to the nonlinear dynamics in Eq.~\eqref{eq:NLDynamics} using MATLAB's \texttt{ode45.m}.

\subsection{Nominal Results with Proposed Station-Keeping Policy}

\begin{figure}[t!]
  \centering
    \begin{subfigure}[]{0.5\textwidth}
       \includegraphics[width=0.99\linewidth]{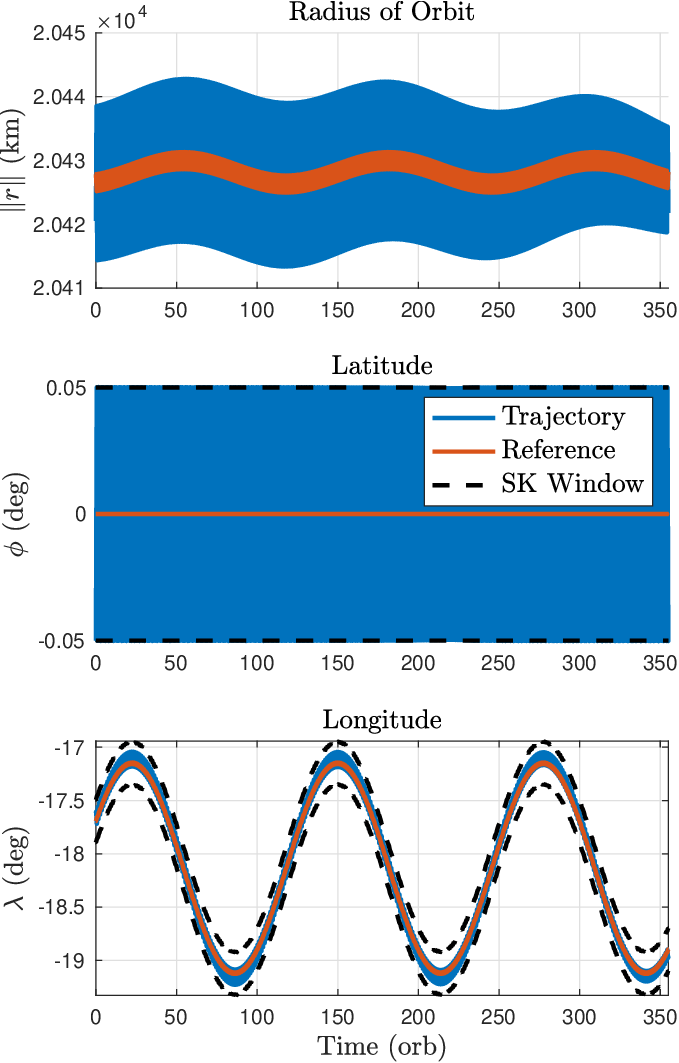}
        \caption{Full one-year simulation results.\label{fig:Inputs_a}}
    \end{subfigure}%
        \begin{subfigure}[]{0.5\textwidth}
        \includegraphics[width=0.99\linewidth]{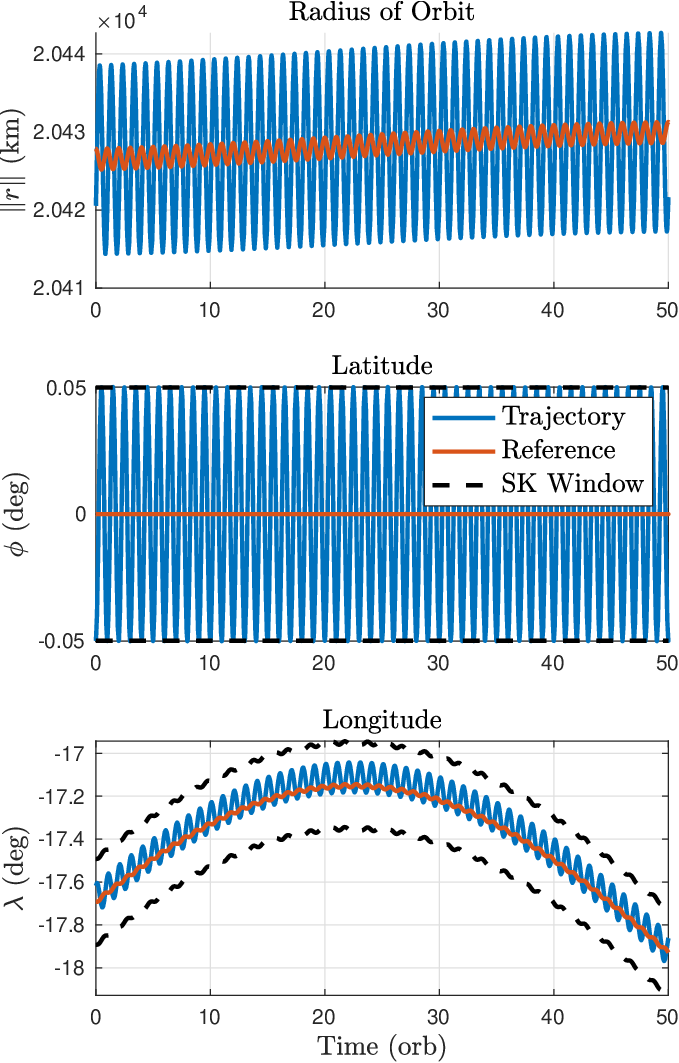}
        \caption{First 50-day simulation results.\label{fig:Inputs_b}}
    \end{subfigure}%
    \caption{Simulated year of the proposed station-keeping policy tracking the natural motion trajectory.}
    \label{fig:Traj}
\end{figure}

\begin{figure}[t!]
    \centering
    \begin{subfigure}[]{0.5\textwidth}
        \includegraphics[width=0.99\linewidth]
        {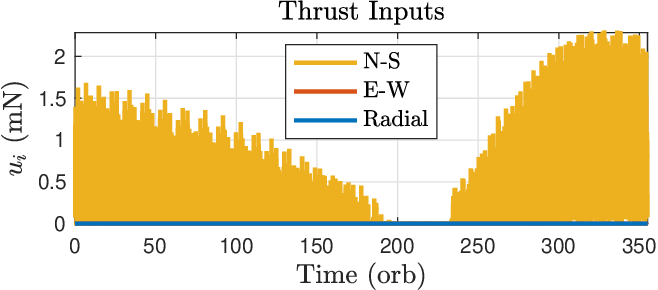}
        \caption{\label{fig:Inputs}}
    \end{subfigure}%
    \begin{subfigure}[]{0.5\textwidth}
        \includegraphics[width=0.99\linewidth]
        {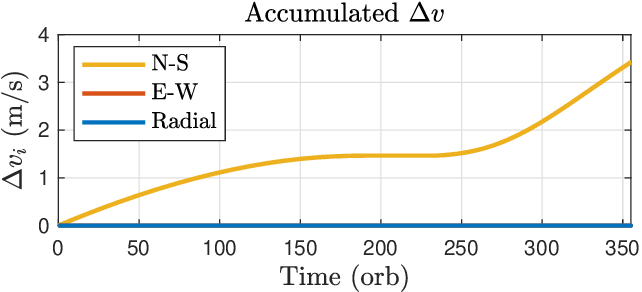}
        \caption{\label{fig:DeltaV}}
    \end{subfigure}%
	\centering
	\caption{Control inputs during a simulated year of the proposed station-keeping policy tracking the natural motion trajectory.
 }\label{fig:Inputs_DeltaV}
\end{figure}

The proposed MPC station-keeping policy is implemented in a simulation of the spacecraft's nonlinear dynamics, where perfect state measurements and knowledge of all perturbations are assumed to be available. The resulting trajectory is provided in Fig.~\ref{fig:Traj}, where the radius, latitude, and longitude of the spacecraft over the entire year is included in Fig.~\ref{fig:Inputs_a} and zoomed in versions of these plots over the first 50 days are included in Fig.~\ref{fig:Inputs_b}. The thrust control inputs associated with this simulation are provided in Fig.~\ref{fig:Inputs} and the accumulated $\Delta v$ is found in Fig.~\ref{fig:DeltaV}. 

The total annual $\Delta v$ achieved by the policy is $3.42$~m/s, effectively all of which is in the North-South direction (the $\Delta v$ expended in the radial and East-West directions is $5.8 \times 10^{-6}$~m/s and $1.7 \times 10^{-6}$, respectively). This highlights the efficiency of the proposed station-keeping policy at mitigating the effect of time-varying and 3rd-body disturbances in the radial and East-West directions. As observed with prior studies, a minimum amount of thrust in the North-South direction is needed to prevent secular drift in the spacecraft's inclination. It is notable in Fig.~\ref{fig:Traj} that the spacecraft does drift within the allowable station-keeping window relative to the natural motion trajectory, likely by taking advantage of the periodic nature of the disturbances acting on the spacecraft to avoid the need to expend fuel.

\subsection{Comparison to Prior AMO Station-Keeping Policies}

To demonstrate the performance of the proposed station-keeping policy relative to methods in the literature, simulations are performed with the LTI-MPC policy from Ref~\cite{Halverson2021}, the LTV-MPC policy from Ref.~\cite{gall2024}, and the nonlinear MPC (nMPC) policy from Refs.~\cite{Halverson2023,Halverson2025}. These policies are implemented within the same simulation setup described in Section~\ref{sec:SimSetup} and with as many of the same tuning parameters used within the proposed policy as possible. The main difference with the policies in the literature is that a static station keeping window is used (i.e., the station-keeping window is defined relative to the nominal AMO orbit at a stable longitude of $17.92^\circ$ West). It is also worth noting that to maintain consistency with Refs.~\cite{Halverson2023,Halverson2025}, the cost function of the nMPC policy is the the 1-norm of the thrust applied, rather than the quadratic cost function used by the other policies. 

The annual $\Delta v$ consumed by all policies is reported in Table~\ref{Table:Comparison}, where the proposed policy has the lowest total $\Delta v$ that is roughly the same as the nonlinear MPC policy from Refs.~\cite{Halverson2023,Halverson2025}.  The LTI-MPC and LTV-MPC policies are the only known convex-optimization-based AMO station-keeping policies in the literature and require $27.3$\% $22.1$\% more total $\Delta v$ than the proposed convex-optimization-based policy. This highlights the contribution of this work, where the proposed policy can be solved as a convex optimization problem, which enables real-time implementation onboard a spacecraft, while achieving similar performance to the computationally-prohibitive nonlinear MPC policy. It is important to note that the high performance and low computation properties of the proposed policy come at the expense of the spacecraft not maintaining a true AMO orbit. In contrast to the LTI-MPC, LTV-MPC, and nMPC AMO station-keeping policies that maintain the spacecraft within $\pm0.2^\circ$ of the stable longitude of $17.92^\circ$ West, the proposed policy maintains the spacecraft within $\pm0.2^\circ$ of the natural motion trajectory, resulting in roughly up to $\pm1.2^\circ$ of longitudinal deviation from $17.92^\circ$ West.

\begin{table}[hbt!]
\caption{\label{Table:Comparison} Comparison of the annual $\Delta v$ required with different station keeping policies.}
\centering
\begin{tabular}{lcccc}
\hline
\hline
Policy & Radial $\Delta v$ (m/s)& E-W $\Delta v$ (m/s)& N-S $\Delta v$ (m/s)& Total $\Delta v$ (m/s)\\\hline
LTI-MPC~\cite{Halverson2021} & 0.161& 0.900 & 3.290& 4.350\\
LTV-MPC~\cite{gall2024} & 0.195& 0.608& 3.371 & 4.175\\
nMPC~\cite{Halverson2023,Halverson2025} & 0.026& 0.088& 3.309& 3.423\\
Proposed Policy and Trajectory & $5.797 \times 10^{-6}$ & $1.724 \times 10^{-6}$ & $3.418$ & $3.418$\\
\hline
\hline
\end{tabular}
\end{table}

\subsection{Robustness Study of Proposed AMO Station-Keeping Policy}

The proposed limit-cycle following AMO station-keeping policy is implemented in realistic off-nominal conditions to assess its robustness properties.  The test cases outlined in this section include uncertainty in the magnitude of thrust provided by the thrusters, uncertainty in the mass of the spacecraft, a lack of knowledge of the time-varying perturbations acting on the spacecraft, a time-delay due to computation time, and the presence of navigation errors.

Thrust uncertainty is injected into the simulation by adding a bias to the thrust commanded by the MPC policy.  This is meant to emulate a thruster with imperfect timing or thrust characterization.  The station keeping policy simulated with a 15\% increase and 15\% decrease in unmodeled thrust results in a 1.4\% increase and 1.9\% increase in annual $\Delta v$, respectively, compared to the nominal results.

The effect of mass uncertainty is assessed by perturbing the mass of the spacecraft's mass to 20\% less than the mass modeled in the MPC policy.  This perturbed simulation results in a 2.6\% increase in annual $\Delta v$ compared to the nominal results.

The nominal results assume accurate knowledge of all perturbation forces acting on spacecraft.  In reality the spacecraft flight computer will not be able to store ephemeris data for all gravitational bodies acting on it.  To assess robustness to this lack of information, simulations are performed without the MPC policy having knowledge of the time-varying gravitational perturbations due to the Martian moons and the Sun.  This results in a 3.8\% increase in annual $\Delta v$ from the nominal results.

The implementation of MPC on a flight computer imposes challenges due to limited on-board computational power and memory.  Prior work claimed that the long time steps associated with MPC-based station keeping policies can mitigate this issue. For example, the 1-hour time steps used in this work may provide substantial time for the relatively simple linear MPC problem to solved, even on a radiation-hardened flight computer.  Although this claim may be sensible, it has not been tested in prior MPC-based station keeping policies.  To better understand this effect, the proposed MPC policy is implemented such that there is a one hour (one time step) delay between the measurement of the spacecraft state and the implementation of the MPC control input.  The resulting annual $\Delta v$ is a 3.9\% increase from the nominal case.

In practice, navigation errors will lead to uncertainty in the spacecraft state used to implement the MPC policy.  To assess robustness to this uncertainty, the proposed station-keeping policy is implemented with zero-mean Gaussian noise added to the spacecraft's position and velocity.  Specifically, a standard deviation of 100~m in position and 0.1~m/s in velocity is used to mimic values that are obtainable in GEO~\cite{Caverly2020}.  This results in a 165\% increase in annual $\Delta v$, with the bulk of this increase occurring due to additional thrust maneuvers in the radial and East-West directions.  Although it is clear that current navigation technology is not capable of such small errors around Mars, this result demonstrates the importance of establishing Mars navigation systems to accurately perform station keeping of spacecraft orbiting Mars.

A summary of the annual $\Delta v$ with all of the robustness case studies is provided in Table~\ref{Table:Robustness}.  The $\Delta v$ accumulated in the radial, East-West (E-W), and North-South (N-S) directions is also provided in this table.  Other than the case with navigation errors, the performance of the proposed station-keeping policy is very robust to realistic modeling errors and computational delays.

\begin{table}[hbt!]
\caption{\label{Table:Robustness} Comparison of the annual $\Delta v$ required with different cases of the proposed station keeping policy.}
\centering
\begin{tabular}{lcccc}
\hline \hline
Case & Radial $\Delta v$ (m/s)& E-W $\Delta v$ (m/s)& N-S $\Delta v$ (m/s)& Total $\Delta v$ (m/s)\\\hline
Nominal  & $5.797 \times 10^{-6}$ & $1.724 \times 10^{-6}$ & $3.418$ & $3.418$\\
15\% Increase in Thrust  & $9.553 \times 10^{-3}$ & $0.0304$ & $3.427$ & $3.467$\\
15\% Decrease in Thrust  & $7.364 \times 10^{-3}$ & $0.0234$ & $3.452$ & $3.482$\\
20\% Decrease in Mass  & $0.0121$ & $0.0393$ & $3.456$ & $3.508$\\
No Knowledge of Time-Varying Perts.  & $2.773 \times 10^{-6}$ & $6.897 \times 10^{-6}$ & $3.549$ & $3.549$\\
1-Hour Computation Delay  &$5.824 \times 10^{-6}$ & $1.739 \times 10^{-5}$ & $3.550$ & $3.550$\\
Navigation Errors  & $1.088$ & $3.650$ & $4.314$ & $9.045$\\
\hline \hline
\end{tabular}
\end{table}

\section{Conclusion}
A limit cycle natural motion trajectory in the longitudinal and radial motion of AMO satellites about a stable longitude was identified. A linear discrete-time model of system dynamics near this natural motion trajectory was constructed and leveraged to enable an efficient predictive station keeping policy. This station keeping policy rivals the performance of nonlinear MPC policies and achieves the lowest $\Delta v$ of any convex AMO station keeping policy in the literature. Moreover, the robustness analysis performed in this work provides an important step towards practical implementation compared to prior AMO station keeping methods in the literature.

\section*{Acknowledgments}
N. A. Gall and R. J. Caverly acknowledge support from a Early Career Faculty grant from NASA’s Space Technology Research Grants Program under award No.~80NSSC23K0075, as well as the University of Minnesota's Research \& Innovation Office and the University of Minnesota's Office of Undergraduate Research. R. D. Halverson acknowledges partial support by the Science, Mathematics, and Research for Transformation (SMART) Scholarship-for-Service Program within the Department of Defense
(DoD), USA.

\bibliography{sample}

\end{document}